# Revisiting Digital Twins: Origins, Fundamentals and Practices

Jiehan Zhou[1], Shouhua Zhang[1,2], Mu Gu[3]

1 University of Oulu (Pentti Kaiteran katu 1, Oulu, 90570, Finland)
2 Heibei University (No.180 Wusi Dong Road, Baoding, 071002, China)
3 Beijing aerospace smart manufacturing technology development Co., Ltd.
Email：jiehan.zhou@ieee.org; shouhua.zhang@oulu.fi; gumu2005@163.com

**Abstract:** The Digital Twins (DT) has quickly become a hot topic since it was proposed. It not only appears in all kinds of commercial propaganda, but also is widely quoted by academic circles. However, there are misstatements and misuse of the term DT in business and academy. This paper revisits Digital Twins and defines it to be a more advanced system/product/service modelling and simulation environment that combines the most modern Information Communication Technology (ICTs) and engineering mechanisms digitization, and characterized by system/product/service life cycle management, physically geometric visualization, real-time sensing and measurement of system operating conditions, predictability of system performance/safety/lifespan, complete engineering mechanisms-based simulations. The idea of Digital Twins originates from modelling and simulation practices of engineering informatization, including Virtual Manufacturing (VM), Model Predictive Control (MPC), and Building Information Model (BIM). Based on the two-element VM model, we propose a three-element model to represent Digital Twins. Digital Twins does not have its own unique technical characteristics; the existing practices of Digital Twins are extensions of the engineering informatization embracing modern ICTs. These insights clarify the origin of Digital Twins and its technical essentials.

**1. Introduction**

With the in-depth integration of computer, information and communication technology (ICT), mechanical engineering, automation and other disciplines, and the unremitting pursuit of human beings for the ultimate performance, such as product innovation, time-to-market, cost, quality, environmental protection, and energy-saving, researchers and practitioners in the industry put forward demands on advanced modelling and simulation environments for computational engineering. Following the introduction of the concept of Virtual Manufacturing (VM) thirty years ago (Onosato and Iwata,1993), its derived term, Digital Twins (DT), was born recently, and frequently cited by business and academic articles.

Gartner had listed DT as one trend of the top ten strategic technology for three consecutive years (2017 - 2019) (Liu et al., 2021). GE has built a DT system of capital flow based on Predix platform, in which engineers and operators can respectively predict the product life cycle (Todorovic et al., 2016). Siemens proposed to use DT to help manufacturing enterprises build an entire production line in the digital space, and digitalize the entire cycle from product design to manufacturing in the physical space (Tao et al., 2018a). In November 2017, the Intelligent Manufacturing Alliance of China officially listed DT as one of the top ten scientific and technological advances in intelligent manufacturing (Wu et al., 2021).

DT has also received extensive attention in academia, and has been introduced into industrial applications. Although major research institutions and related enterprises have presented their own DT concepts, there are many different definitions of DT since it was proposed (Opoku et al., 2021). With the continuous interpretation of DT in industry and academia, its meaning has become more confusing, and the boundaries between DT and other related concepts have become more obscure. What exactly is a DT, what can it do, where is the boundary, what is its relationship with modelling and simulation? These questions are confusing researchers and practitioners. Because of the unclear definition of DT, its technology is ambiguous, and it is even more ambiguous under the entertainment of the nascent metaverse. Numerous researchers across the fields of computer, manufacturing, and automation are overwhelmed by the influx of literature terminology. Even more notable phenomena in academy are that many researchers just added DT in the title of the articles based on existing traditional engineering informatization to gain popularity (Zhang, 2020).

In response to the misuse and abuse of the term DT, this paper profoundly traces the source of DT by consulting a large number of first-hand materials, corrects such claim that Digital Twins originated from the mirrored space model (MSM), and corrects the misstatement that DT first appeared in the technical report on Modelling, Simulation, Information Technology and Processing Roadmap by National Aeronautics and Space Administration (NASA) in 2010, and defines the basic concept of DT as a more Advanced System Modeling and Simulation Environment (ASM&SE). The DT-based system has the characteristics of not only geometric visualization of the physical system, but also real-time sensing and measurement of system operating conditions, the predictability of system performance/safety/lifespan, the complete engineering mechanisms-based simulation, the symbiosis of physical/virtual systems, and so on. The terms such as system, product, and service are used interchangeably in the paper. Based on this novel understanding, and the two-element model of Virtual Manufacturing, we present a three-element model of DT, namely the geometric shape of a real system, the information of the real system, and the engineering mechanism of the real system. Based on the analysis of DT practices, the paper pointed out that the current industrial application of DT is essentially an informatization solution that combines the engineering requirements, and expands the application of modern ICT.

The remainder of the paper is organized as follows. Section 2 traces the source of the DT idea. Section 3 defines DT as an advanced system modelling and simulation environment based on a digital gearbox product life cycle management case analysis. Section 4 reviews the two-element model for representing VM and proposes a three-element model for representing DT. Section 5 briefly describes some practical research in the name of DT, and points out that current practices are an extension of traditional engineering informatization, which partially reflects the characteristics of DT, such as the geometric visualization of a physical system. Section 6 draws a conclusion.

**2. The origin of DT**

The idea of DT is essentially an ASM&SE. The current widely circulated version is that DT originated from MSM coined by Michael Grieves at the Florida Institute of Technology (Tao et al., 2018b; VanDerHorn and Mahadevan, 2021). This MSM model was first mentioned in a report by Grieves in late 2002 when he was studying virtual-real-driven product life management (PLM). The term MSM was used in the first PLM courses at the University of Michigan in 2003 (Kritzinger et al., 2018). In 2005, the MSM term appeared in the article "Product life cycle management: The new paradigm for enterprises" (Grieves, 2005); In 2006, the MSM term was changed to the Information Mirror Model (IMM) (Grieves, 2006). In the book "Transdisciplinary Perspectives on Complex Systems" edited by Kahlen et al., Grieves and Vicker co-published an article "Digital Twin: mitigating unpredictable, undesirable emergent behaviour in complex systems" (Grieves and Vickers, 2017). In the paper, Grieves stated that DT originated from his MSM, then NASA's Vickers and others borrowed his MSM idea and applied it to NASA's technical report in 2010. Actually, his claim in 2017 was not mentioned in NASA's technical report published in 2010, and there is no conclusive documentary or photographic evidence that DT originated from Grieves' MSM.

In fact, the team led by Prof. Iwata at Osaka University proposed VM as the "virtual manufacturing modelling and simulation environment" and developed its prototype as early as 1993 from the perspective of the original idea of DT on " the virtual representation of a physical system". The concept of VM is nearly 10 years earlier than MSM, and has a history of 30 years. VM combines concepts such as real physical system (RPS), virtual physical system (VPS), real information system (RIS) and virtual information system (VIS), and identifies four categories of VM systems based on ICT. Among the numerous DT review papers, Iwata's VM is gradually attracting the eyes of a few scholars. For example, Semeraro et al. cited and explicitly mentioned the contribution of Iwata's team in their review article "digital twin paradigm" (Semeraro et al., 2021): virtual manufacturing is defined as a system aimed at generating a virtual representation of a physical system without using real facilities/entities (Onosato and Iwata, 1993).

Second, the industry generally believes that the term DT first appeared in the Modelling, Simulation, Information Technology, and Processing Roadmap published by NASA in 2010 (Deng et al., 2021). This statement is also inconsistent with the fact. In fact, the term DT first appeared in the article written by Hernández et al. in 1997 (Hernández et al., 1997). Although Hernández et al. did not define DT in the paper, it can be considered semantically that DT is a three-dimensional digital model of the urban road network, and it is expected that the model can be iteratively modified and coexist with the physical system in implementation. The DT, which appeared in NASA's technical report 13 years later, aims to explore the digitization of advanced manufacturing, integrate and drive modern aircraft design, manufacturing, operation and maintenance, real time cope with the complexity from hardware operation and maintenance, and expects to significantly reduce product cost and time to market. It emphasizes presenting techniques that can digitalize multidisciplinary physical models that not only characterize physical materials, but also how a system operates. These models can be used in the production and operation of spacecraft (Shafto et al., 2010).

Two other concepts that have contributed to the idea of DT are model predictive control (MPC) and building information model (BIM). The core idea of MPC is that applying models in each control cycle predicts the dynamic characteristics of a system, and then seeks the finite-time open-loop optimal control strategy in the current control cycle (Garcia et al., 1989). DT and MPC simulate current states to predict future conditions, but the goal of DT is to create virtual models in synchronization with their physical systems.

BIM keeps accurate and interoperable records of building information to enhance planning, construction and maintenance over the life of a facility (Khajavi et al., 2019). The main difference between architectural BIM and DT is that BIM is designed to increase design and construction efficiency, not real-time data. DT leverages real-time data to simulate and control physical systems, improve operational efficiency, and enable predictions. Table 1 presents the timeline of DT ideas and terminology.

**Table 1**. Timeline of DT ideas and terminology

|  | Earliest publication | Misinformation or self-reporting |
|---|---|---|
| Source of DT | VM, 1993 | MSM, 2003 |
|  | MPC 1970's | NASA, 2010 |
|  | BIM 2000's |  |
| Earliest reported DT | Hernández et al.,1997 | NASA, 2010 |

It is safe to say that the idea of DT has been bred since humankind developed and used computers. The earliest and most persuasive research on DT can be traced back to the VM system modelling and simulation environment proposed by Iwata et al. in 1993. Their idea on VM is still applicable as a system modelling and simulation environment as targeted by its derived version, i.e., DT. In conclusion, either early VM or its' derived DT, their theoretical development and engineering practice deeply depend on the in-depth development and integration of interdisciplinary engineering knowledge informatization, industrial software, and ICT.

## 3. Case study and definition of DT

This section introduces DT by taking an example of the modelling and simulation of a gearbox. Assuming that a gearbox consists of a pair of gears A and B. In the design stage, a designer applies computer-aided design (CAD) to produce a digital copy of the gearbox according to the design requirements. Before the actual manufacturing, computer-aided engineering (CAE) is adopted to simulate and analyze the structure/strength/wear/lifespan performance of the gear pair. The analysis results are fed back to the design stage. The designer optimizes and corrects the digital copy; the designs with satisfied simulation results enter the manufacturing/assembly stage. Before the actual manufacturing/assembly, it is also necessary to use computer-aided manufacturing/assembly software to simulate the machining and assembly process of the gear pair. During the actual operation and maintenance of the gearbox, the staff applies the embedded software and sensor systems to collect various physical parameters generated by the running gearbox in real time, such as temperature, rotational speed, and stress/strain, and applies these data to create a digital 3D gearbox model in computer world corresponding to the gearbox in the real world. This re-constructed digital gearbox model (i.e., gearbox twin) can help operators visually observe and predict system abnormalities and send out maintenance and control commands in time.

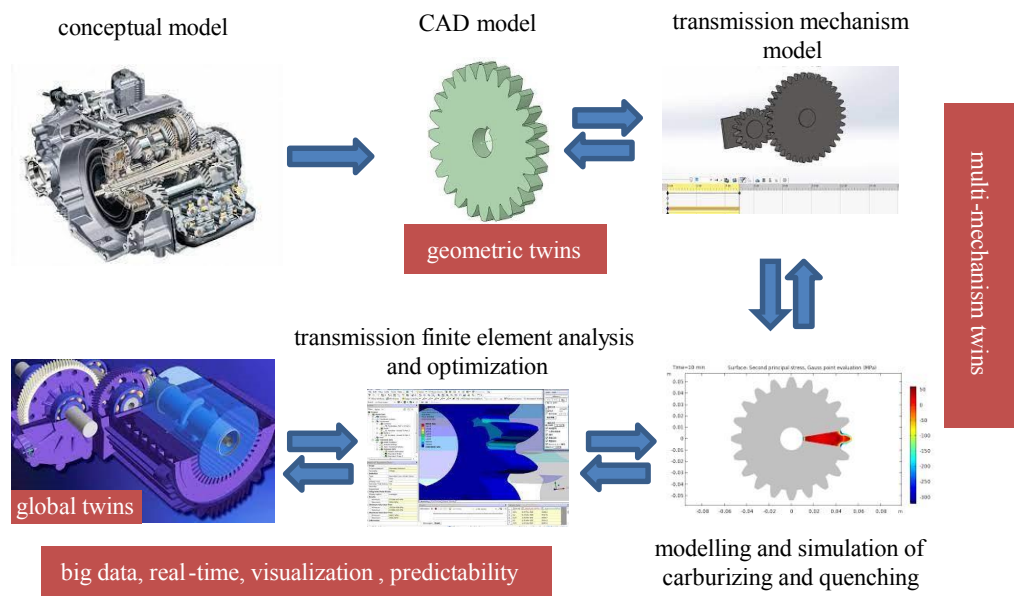

**Fig 1**. The schematic diagram for the digitization of a gearbox life cycle support by DT modelling and simulation environment

The above case study briefly describes the digitization of a gearbox from the conceptual phase to the manufacturing, operation and maintenance phases (Figure 1), during which various computer-aided tools such as CAD, CAE, computer-aided manufacturing (CAM), and embedded real time monitoring systems are deployed for gearbox design, performance simulation, operation, and maintenance. Following this case study, we will further clarify what is DT by answering the following questions:

Can we call the computational geometric model (CAD) of the gearbox and gear pair the DT of the gearbox and gear pair? Our answer is yes; it is their geometric digital twins.

Give the gears a certain number of teeth and rotational speeds for a computer motion simulation. Can we call the computational motion simulation the DT of the gears? Our answer is yes; it is their physically motional digital twins.

Mesh the gear pairs A and B for a relative motion simulation based on the gear transmission mechanism in the computer, and can we call the gear transmission simulation the DT of the gear pair? Our answer is yes; it is their physical transmission mechanism digital twins.

There are a series of computational collision/meshing simulations evaluating the contact/stress/strain/deformation of the gear pair with various speeds and stiffness. Can we call meshing collision simulation the DT of the gear pair? Our answer is yes; it is their contact/collision engineering mechanism digital twins.

In order to enhance the wearing lifespan of the gear surface, we need to operate the coating simulation of gear as well. Can we call the carburizing simulation the DT of the gear pair? Our answer is yes; it is their carburizing/stiffness/materials engineering mechanism digital twins.

Moreover, the global digital gearbox connects and communicates with the running physical gearbox in the operation and maintenance phases. The static and dynamic characteristics of the real physical gearbox are one-to-one mapped to the computer world, such as geometric models, various information on temperatures, and speeds, and various engineering mechanisms in real time. Can we call this global digitization the DT of the gearbox? Our answer is yes; it is the global digital twins.

We can see that it is more reasonable to define DT as an advanced computational modelling and simulation environment for system/product/service life cycle management. This computational environment supports the life cycle management of products/services/systems, not only from the digitization of conceptual design, computational mechanisms simulation, but also the real-time operation and maintenance with possible geometric mirroring. DT comprehensively applies modern ICTs, industrial software, and engineering knowledge such as computational geometry, computational engineering, and Internet of Things (IoT) for collecting and transmitting data on system operating conditions in real time, computational fluid dynamics (CFD) for simulating fluid mechanisms, and artificial intelligence for predicting system health. DT-transformed industrial application has the following characteristics:

1) System/product/service life cycle management support. DT emphasizes building a modelling and simulation environment for the entire system life cycle management in the computer world, rather than only for the modelling and simulation of the global system in operation and maintenance stage.

2) 3D geometric modelling of real systems. Hernández et al. (1997) first used the term DT, regarded DT as a 3D digital model of an urban transportation network. This characteristic is generally acknowledged by most works, including Grieves' MSM (Kritzinger et al.,2018), but it should not be the essential feature of DT.

3) The real-time sensing and measurement of system operating conditions. With the development and penetration of the Internet of Things in industrial applications, DT can collect and transmit data and information on system operating conditions in real time, and provide big data for the subsequent various mechanisms simulations and system performance predictions.

4) Multi-mechanism modelling and simulation. Mechanisms modelling (from data to model) is the digitization of engineering knowledge with a result of industrial software such as the solver for computational flow dynamics. Mechanisms simulation is the inverse modelling process (from model to data) for system functional and non-functional evaluations. The system modelling and simulation under the DT paradigm emphasizes the modelling and simulation for the system complexity and multiple mechanisms, such as simulations on materials, geometric structures, structural strength, system lifespan, etc.

5) Proactive system performance prediction. DT makes full use of real-time big data, employs machine learning, deep learning, and distributed software and hardware architecture, such as cloud computing, to predict system performance, such as equipment failures, load balance, system lifespan, and so on.

6) Real-time system simulation and control. In addition to supporting system simulation, another core goal of DT is to perform real-time simulation for system operation and maintenance, and send real-time control commands for extending system lifespan.

Leng et al. (2021) presented the enabling technology map for DT. It can be seen that DT itself is not like the concepts of Virtual Machine and Cloud Computing with distinct technical characteristics, but is a collective term with two major technology clusters of modern ICTs and computational engineering know-how (i.e., digitalized engineering knowledge/industrial software). These modern ICTs include industrial IoT, real-time synchronization, discrete event simulation, visualization, big data analytics, industrial artificial intelligence, industrial blockchains, cloud computing, and so on. The computational engineering know-how digitalizes engineering knowledge with CFD, finite element analysis, etc., as result (Figure 2).

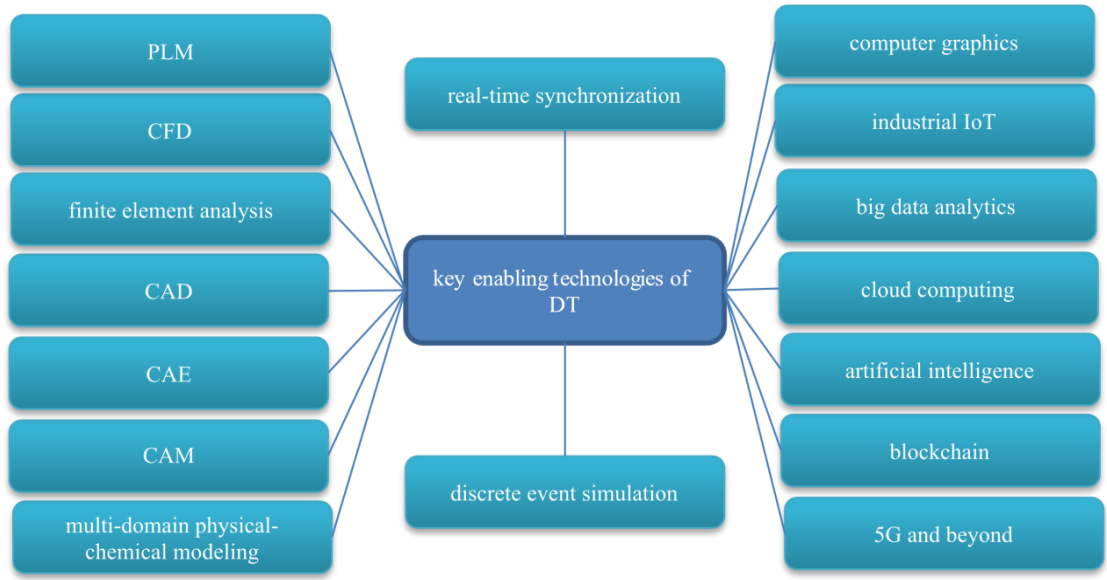

**Figure 2**. Key enabling technologies for DT

**4. Three-element DT model**

As mentioned in the review by Semeraro et al. (2021), the idea of DT mainly originates from VM, MPC, and BIM. This section focuses on reviewing the two-element model of VM proposed by Iwata et al. (1993), and proposes an extended version, a three-element model of DT. Readers who are interested in MPC and BIM can refer to the literature (Qin and Badgwell, 2003; Garcia et al., 1989; Volk et al., 2014; Miettinen and Paavola, 2014).

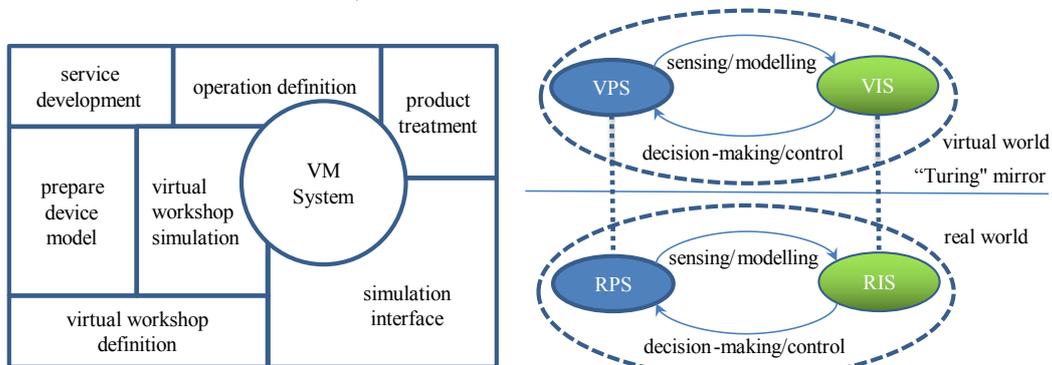

Figure 3. The two-element model of (VM): RPS (real physical system), RIS (real information system), VPS (virtual physical system), VIS (virtual information system)

Figure 3 (left) presents the modelling and simulation architecture of a VM system composed of 7 functional modules (Iwata et al., 1993; Zhou et al., 2000). Figure 3 (right) presents the core idea of the two-element model of VM as follows: a real manufacturing system consists of RPS and RIS, a VM system consists of VPS and VIS. VPS and RPS have similarities in geometric structures and logical functions, and VIS and RIS are equivalent in the amount of information. It can be seen that this two-element model emphasizes more computational modelling and simulation of a real manufacturing system, that is, a real physical system and a computerized system can interact in information.

The two-element model assumes that the real world is composed of RPS and RIS, which ignores engineering knowledge or generally includes engineering knowledge into the information system. Here, we divide the information system in the two-element model into two parts: information and engineering knowledge/know-how (i.e., engineering mechanisms), and propose a three-element model of DT (Figure 4). We believe that a real world can be represented with three elements: real physical system (shape), real information system and real mechanism system. The purpose of building a DT is to use computer to approximately represent the three elements in the real world, such as using computer graphics to generate the 3D geometric model of a real scene, and using CFD to simulate the motion performance of a real aircraft. The three-element model is also in line with the gradual human's cognitive processes from shapes, information to mechanisms of a real system.

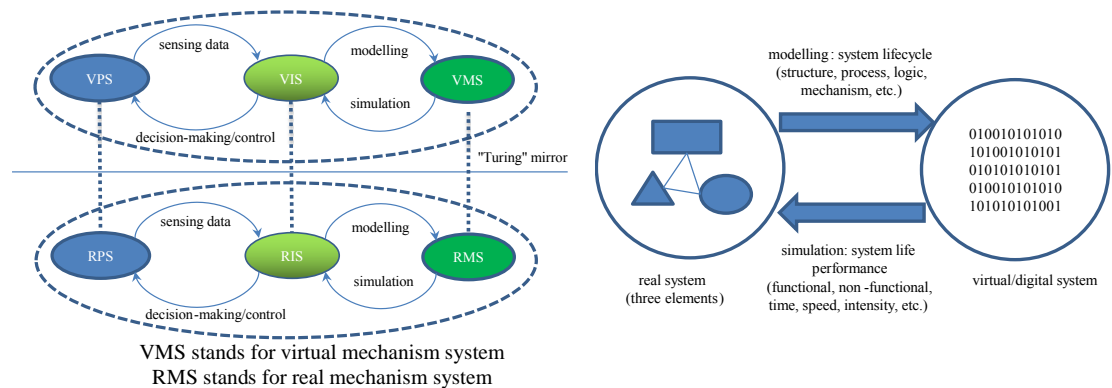

VMS stands for virtual mechanism system
RMS stands for real mechanism system

**Figure 4**. Three-element model of DT

## 5. Review and analysis of DT practices

With the rapid development of geometric modelling and simulation, data perception, high-performance computing, and high-speed wireless communications, the concept of DT has been gradually applied to engineering practices, and implemented at the system operation and maintenance level. However, based on our definition of DT as ASM&SE, the following DT practices are largely the extension of traditional informatization and digitization, or the informatization solution to an engineering problem in a professional field. They might support some of the given DT features by this paper, such as geometric visualization, real-time data collection, predictability of operating conditions with deep learning, etc.

### 5.1 Smart manufacturing

Ghosh et al. proposed the concept of twin based on sensor signals, developed a DT construction system (DTCS) and DT adaptation system (DTAS) on a JavaTM-based platform, and used real-time processing of milling torque signals as an application case (Ghosh et al., 2021). DTCS constructs DT based on a delay-embedded signal processing method. DTAS adapts constructed DT. DTCS consists of five modules: input, modelling, simulation, validation, and output. DTAS only uses simulated signal datasets that the validation module tests positive while monitoring progress. It receives real-time signals from the machine tool for monitoring purposes. Any update in DTCS will change the content. DTAS will also update itself and confirm the changes in DTCS in real time, which makes the two systems highly coupled.

This DT practice covers RPS since sensor signal delays are real in intelligent machine tools. It covers RIS because the simulated and real delayed signals are compared to determine the course of action. This DT practice covers VPS and VIS by modelling and simulating delayed signals from sensors signals. This

DT practice covers RMS and VMS because the mechanism of delayed signals is applied when modelling and simulating delayed signals.

Ghosh et al. pointed out that the goal of DT is to provide a computer system that helps build and use twins, which is consistent with the idea of VM modelling and simulation environment. The studied case is a real-time monitoring of machine tools based on embedded sensor networks. Compared with the traditional monitoring system, it emphasizes the feature of real-time modelling and simulation.

**5.2 Smart building**

Khajavi et al. proposed a method for establishing a sensor network to create the DT of a building (Khajavi et al. , 2019). This is achieved by collecting and analyzing specific environmental factors in the exact surrounding of the building in real time. Although this study utilizes only a limited sensor network and three environmental parameters for sensing (i.e., light, temperature, and humidity), the introduced step-by-step framework can be used to create a more comprehensive DT of a building facade and a building interior.

This DT practice constructs a geometric model of a building façade and presents the light, temperature, and humidity of the building façade in real time. Therefore, it covers RPS and VPS. In real life, we can adjust equipment such as lighting and air conditioners according to the environmental parameters. Therefore, it covers RIS and VIS. Since it has no further processing and utilization of the environmental parameters, it does not cover RMS and VMS. Compared to BIM, it emphasizes the real-time requirement.

**5.3 Smart energy**

Singh et al. (2021) presented a toolbox for implementing DT to enhance modelling and simulation practices. The toolbox can realize the DT of a battery system for a micro-robot vehicle. They reviewed DT from the perspective of modelling and simulation, and proposed the implementation method, including the DT framework and process model, and gave a case analysis of a battery system. The battery DT can be used to estimate the state of health of a battery and optimize battery life by evaluating the capacity fading with the number of cycles. Compared to the existing tools for implementing DT models, this approach focuses on defining the required features of a DT model and then selecting the relevant tools.

This DT practice does not construct a geometric model for the battery. Therefore, it does cover RPS, but not VPS. It covers RIS and VIS due to health monitoring based on real-time collected battery data. It estimates capacity decay using an extended Kalman filter and uses an equivalent circuit model to simulate the electrical behavior of a battery, so it covers RMS and VMS.

**5.4 Smart agriculture**

Pylianidis et al. (2021) analyzed the value-added services of DT for agriculture, and gave the development route of agricultural DT. They believe that there is no unified definition of DT for various disciplines. A feasible definition emphasizes a virtual representation of a dynamic physical object/system, which spans multiple life cycle stages and provides decision-making with the use of data analysis methods. The so-called agricultural DT is another term for Agricultural Informatization or Smart Agriculture with the help of new ICT technologies. Most current DT practices in Smart Agriculture cover RPS, RIS and VIS.

There are other practices, such as the marking robot DT prototype exhibited in Hannover Messe 2018 (Figure 5). The exhibition showed that the robot/arm in the real world perform marking tasks such as lifting, rotating, etc., which are digitally mapped to the computer world in real time, accompanied by the real-time visualization of multiple physical signals. Based on the rapid scene constructing, design model optimization, and configuration simulation, this DT re-constructs a digital marking robot, and realizes the mapping of a physical marking production line to a virtual world based on MODBUS (Modbus, 2022)

TCP protocol. the marking robot DT demonstrates that it can dynamically make plan scheduling and operation optimization through transmiting the optimized parameters back to the physical system in real time. Therefore, this marking robot DT covers RIS, RPS, RMS, VIS, VPS and VMS. Table 2 summarizes the reviewed DT practices from the perspectives of the three-element DT model.

As presented, there are several major features of DT applications: product/service/system life cycle management, geometric visualization of physical shape, real-time operating conditions sensing and measurement, real-time mechanisms modelling and simulation, predictability of system health. But it is not necessary for a DT application to present all these features one time. We define DT as an ASM&SE, and its goal is to provide digitization methods and techniques to support these features. Compared with the traditional real-time marking robot, the DT system (Figure 5) added the visualization of the 3D robot model. This addition may have value in the safety evaluation of human-robot interaction. However, this addition may not be necessary in the case of robot failures prediction. On the contrary, it will increase the development costs and difficulties. Therefore, applying DT needs to consider its necessity and build the right digital twin (Zhang et al. ,2021).

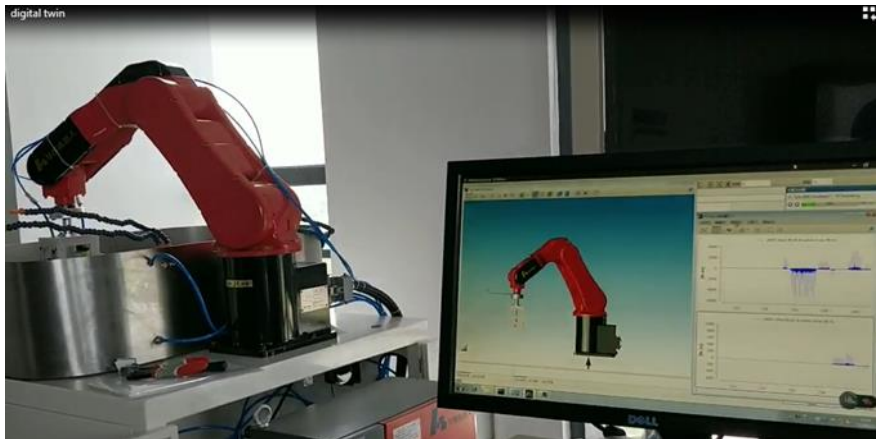

**Figure 5**. Marking robot DT prototype exhibited in Hannover Messe 2018

**Table 2**. DT practices

| Application | RPS | RIS | RMS | VPS | VIS | VMS |
|---|---|---|---|---|---|---|
| Ghosh et al., 2021 | √ | √ | √ | √ | √ | √ |
| Khajavi et al. ,2019 | √ | √ |  | √ |  |  |
| Singh et al., 2021 | √ | √ | √ |  | √ | √ |
| Smart agriculture | √ | √ |  |  | √ |  |
| The marking robot DT | √ | √ | √ | √ | √ | √ |

## 6. Conclusion

There have been so many different opinions about DT and its relevant practices. However, there are misstatements and misuse of DT as well. This paper revisits the origin of the DT idea from the perspective of an advanced modelling and simulation environment. According to literature records, Virtual Manufacturing and its practice proposed by Iwata et al. in 1993 is the earliest research and practice on Digital Twins to our knowledge. DT is an ASM&SE that deeply embraces contemporary ICTs and computational engineering knowledge with software to re-construct the three elements of a product/service/system: physical shape, physical information, and engineering mechanisms. The ultimate goal of DT is to achieve the transparent and predictable operation and maintenance of a real physical system. A DT application is characterized by the geometric visualization of a real system, real-time data collection of its operating conditions, complete engineering mechanisms modelling and simulation, and health prediction. Modern ICTs and the digitization of engineering knowledge are the fundamental enabling technology of DT. Clearly, the existing DT practices are trying to apply or extend modern ICTs into proposing an informatization solution to a professional problem.

For DT researchers and practitioners in engineering, the cornerstone of DT relies on engineering mechanisms modelling and simulation. Reflected in manufacturing, it is the research and development (R&D) of modelling and simulation tools for engineering knowledge. The results are collectively referred to as industrial software, such as CAD, CAE, CAM, and CFD. With the rapid development of cloud computing, industrial software as a cloud service is gaining widespread attention from academia and industry. Recently, traditional giant industrial software companies claimed to be able to provide customers with DT solutions, for example, CATIA 3D Experience Virtual Twin by Dassault (2022), Fusion 360 by Autodesk (2022). Essentially, they are selling similar modelling and simulation services with the label of Digital Twins.

For DT researchers and practitioners in ICTs, the cornerstone of DT relies on advanced ICTs such as computer graphics, artificial intelligence, Internet of Things, 5G and beyond, blockchain, mixed reality, and mathematics to address high-performance 3D reconstruction, rendering and simulation, real-time data sensing, predictable system maintenance, high-speed data communications, transaction security, unbounded interaction virtual and real world, etc.

**Acknowledgments**

Thanks to Dr. Haibin Yang from Wuhan Huazhong Numerical Control Co., Ltd. and Dr. Xudong Cai from CASICould for sharing the demo exhibited in Hannover Messe 2018.